\documentstyle[aps,prl,multicol,epsf]{revtex} 
\begin{document} 
\widetext
\title{\bf Interrelation between the pseudogap and the 
incoherent quasi-particle features of high-$T_c$ superconductors}

\author{J. ~Ranninger$^{(a)}$ and 
A.~Romano$^{(b)}$}

\address{$^{(a)}$ Centre de
Recherches sur les Tr\`es Basses Temp\'eratures, 
Laboratoire
Associ\'e \'a l'Universit\'e Joseph Fourier, 
\\ Centre National de la
Recherche Scientifique, BP 166, 38042, Grenoble 
C\'edex 9, France}

\address{$^{(b)}$ Dipartimento di Scienze Fisiche "E.R. Caianiello", 
Universit\`a di Salerno, I-84081 Baronissi (Salerno), Italy\\
Unit\`a I.N.F.M. di Salerno}

\date{March 31, 1998} 
\maketitle 
\draft 
\begin{abstract}
Using a scenario of a hybridized mixture of localized bipolarons and 
conduction electrons, we demonstrate for the latter the simultaneous 
appearance of a pseudogap and of
strong incoherent contributions to their quasi-particle spectrum 
which arise from phonon shake-off effects. This can be traced back 
to temporarily fluctuating local lattice deformations, giving rise to a 
double-peak structure in the pair distribution function, which should 
be a key feature in testing the origin of these incoherent 
contributions, recently seen in angle resolved photoemission 
spectroscopy ($ARPES$). 
\end{abstract}


\begin{multicols}{2}

\narrowtext

The appearance of a pseudogap, accompanied by a predominantly 
incoherent quasi-particle spectrum  
in certain parts of the Brillouin zone\cite{ARPES}, 
is considered to be amongst  the most significant signatures of 
high-$T_c$ superconductors ($HT_cSC$) which may 
contain the key of our understanding of these materials. As suggested 
earlier, the large incoherent part of the 
quasi-particle spectrum might come from a coupling of the electrons to 
collective modes such as spin fluctuations\cite{Schrieffer-97}. We 
shall discuss and defend in this Letter a similar point of view based 
on a scenario of a mixture of intrinsically localized bipolarons and 
coexisting itinerant electrons, hybridized with each other via  charge 
exchange, permitting bipolarons to disintegrate into pairs of 
conduction electrons and vice-versa to reconstitute themselves in an 
inverse process. The location of the bipolarons in high-$T_c$ materials 
might be sought in the highly polarizable dielectric layers adjacent 
to the $CuO_2$ planes or possibly inside the polaronic 
stripes\cite{Bianconi-97} in those planes themselves - the 
remainder of those $CuO_2$ planes forming the subsystem 
housing the itinerant electrons. Taking the bipolarons as 
quasi-particles without any internal structure,
such a scenario is described by the so called Boson-Fermion 
model ($BFM$) which has led us to the prediction of  
pseudogap features in the quasi-particle spectrum\cite{Ranninger-95}, 
driven by strong local electron-pair correlations. In the present 
Letter we extend our previous studies by taking into account the 
internal polaronic structure of the bipolaronic Bosons, as being composed 
of charge and lattice vibrational degrees of freedom, 
locked together in a coherent quantum state. 
A bipolaronic Boson localized on a site $i$ is represented by 
\begin{equation}
b^{+}_i~e^{-\alpha(a_i-a^{+}_i)}|0\rangle | \Phi(X) ) =  
b^{+}_i|0\rangle | \Phi(X-X_0) ).   \quad  \label{equ1}
\end{equation}
where the phonon operators $a_i^{(+)}$ correspond to local lattice 
deformations. The hard-core Bose operators $b_i^{(+)}$ describe 
pairs of electrons which are self-trapped inside locally deformed
clusters of atoms, characterized by deformed harmonic oscillator 
states $|\Phi(X-X_0))$ with equilibrium positions shifted by 
$X_0=2\alpha\sqrt{\hbar/2M\omega_0}$
($\omega_0$ denotes the characteristic frequency and 
$M$ the mass of the oscillators). 
The strength of the coupling of the charge 
carriers to local lattice deformations, ultimately leading 
to bipolaron formation, is given by $\hbar \omega_0 \alpha$.
Such physics is described in terms of the following generalization
of the original $BFM$: 
\begin{eqnarray}
H & = & (D-\mu)\sum_{i,\sigma}c^+_{i\sigma}c_{i\sigma}
-t\sum_{\langle i\neq j\rangle,\sigma}c^+_{i\sigma}c_{j\sigma}
\nonumber \\
& + & (\Delta_B-2\mu) \sum_ib^+_ib_i
+v\sum_i [b^+_ic_{i\downarrow}c_{i\uparrow}
+c^+_{i\uparrow}c^+_{i\downarrow}b_i] \nonumber \\
& - & \hbar \omega_0 \alpha \sum_ib^+_ib_i(a_i+a_i^{+})
+\hbar \omega_0 \sum_i \left(a^{+}_i a_i +\frac{1}{2}\right). 
\label{eq2}
\end{eqnarray}
Here $c_{i\sigma}^{(+)}$ are Fermionic operators referring to  
itinerant electrons with spin 
$\sigma$. The bare hopping integral for 
the electrons is given by $t$, the bare Fermionic half band 
width by $D$, the Boson energy level by 
$\Delta_B$ and the Boson-Fermion pair-exchange coupling 
constant by $v$. The chemical potential $\mu$ is common to  
Fermions and Bosons.
The indices $i$ denote effective 
sites involving molecular units made out of  
adjacent molecular clusters of the metallic Fermionic and 
dielectric Bosonic subsystems. Because of the small overlap of 
the oscillator wave functions at different sites we may, to 
within a first approximation, consider the Boson and Fermion 
operators as commuting with each other. 

The original $BFM$, given by the first two lines in 
Eq.(2), has been investigated in great detail 
as far as the opening of the pseudogap is concerned and as far as 
this affects the thermodynamic, 
transport and magnetic properties\cite{Ranninger-95}. The opening of 
the pseudogap in the Fermionic density of states was shown to be 
driven by the onset of local electron pairing without any 
superconducting long range order. Even without treating the 
generalized $BFM$ within the self-consistent conserving approximation, 
used in those studies of the original $BFM$, we find that 
the atomic limit of this generalized $BFM$ already gives us clear 
indications on the interrelation between the opening of the pseudogap 
and the appearance of predominantly incoherent quasi-particle features,
as seen in ARPES studies.

In order to set the scale of the various parameters in this model 
we measure them in units of 
$D$, which for typical $HT_cSC$ is of the order of $0.5~eV$. 
As in our previous calculations of the original $BFM$, we choose $v$ 
such that the pseudogap opens up at temperatures of the order 
of a hundred degrees $K$. We take  $v=0.25$ for the present 
study. We furthermore choose $\alpha=2.5$ such that together 
with a typical local phonon frequency  of the order of 
$\omega_0=0.1$ we have a reasonable bipolaron binding energy 
$\varepsilon_{BP}=\alpha^2 \hbar \omega_0$ which pins the chemical 
potential at about half the renormalized Bosonic level 
$\tilde\Delta_B= \Delta_B- \hbar \omega_0 \alpha^2$. We choose 
$\tilde\Delta_B$ to lie close to the band center such that the 
number of electrons is slightly below half-filling (typically 
around $0.75$ per site, which is the physically relevant regime 
of concentrations). For larger binding energies the bipolaronic level 
would drop below the band of the electrons leading to a situation of 
{\it bipolaronic superconductivity}, which is clearly not 
realized in $HT_cSC$ since they definitely show a Fermi surface.

The idea behind applying the Boson-Fermion scenario to $HT_cSC$ is 
that we are confronted with inhomogeneous systems consisting of highly 
polarizable substructures on which localized bipolarons are formed. 
These local substructures are embedded in the rest of the 
lattice\cite{Roehler-97} which is occupied by electrons having 
a large either hole or electron-like Fermi 
surface\cite{Ding-97,Ino-98}, depending on doping. 
In such a two-component scenario
the electrons scatter in a resonant fashion in and out of the Bosonic 
bipolaronic states. It is that which is at the origin of the 
opening of the pseudogap in the normal state 
of these materials, driven by a precursor of electron 
pairing\cite{Ranninger-95}, rather than magnetic interactions
\cite{Ding-97}. Generalizing this scenario in the way 
described above provides a mechanism by which the electrons 
acquire polaronic features (which, unlike for Bosons, are not of 
intrinsic nature) via the charge exchange term. This term thus 
not only controls the opening of the pseudogap as in the original 
$BFM$ but also the appearance of the strong incoherent contributions 
to the electron spectrum arising from phonon shake-off effects. 
Given the two-subsystem picture on which the Boson-Fermion model 
is based, doping leads primarily to the creation of localized 
bipolarons which beyond a certain critical concentration are 
exchanged with the itinerant electrons. For a system such as, for 
instance, $YBCO$ the number $n_B=\langle b_i^+ b_i \rangle $  
of doping induced bipolarons (approximately given by half the 
number of dopant $O_2^{2-}(1)$ ions in the chains) varies between 
$0$ and $0.5$ per effective site and the number of Fermions  
$n_F=\sum_{\sigma} \langle c^{+}_{i\sigma}c_{i\sigma} \rangle$ is
equal to 1
if the Boson-Fermion exchange coupling were absent. We thus obtain 
a total number of charge carriers $n_{tot}=n_F+2n_B$ close to 2 for 
optimally doped systems. We should however remember that in real 
systems doping not only changes $n_{tot}$ but also the relative  
occupancy of Fermions and Bosons, which seems to be the most 
important effect in the doping mechanism of these materials, 
achieved in our model as soon as $v$ is different from zero. 

We shall in the following solve the generalized $BFM$ in the 
atomic limit (i.e., putting the second term in Eq.(2) equal to 
zero) for a grand canonical ensemble. In this case the eigenstates 
of the Hamiltonian are
\begin{eqnarray}
|0,l \rangle& =& |~0\rangle \otimes |0) \otimes |\Phi(X)\rangle_l 
\nonumber \\ 
|1,l \rangle& =& |\uparrow\rangle \otimes |0) \otimes 
|\Phi(X)\rangle_l 
\nonumber \\
|2,l \rangle& =& |\downarrow\rangle \otimes |0) \otimes 
|\Phi(X)\rangle_l 
\nonumber \\
|3,l \rangle& =& u_{l,+}|\uparrow \downarrow \rangle \otimes |0) 
\otimes |\Phi(X)\rangle_{u_{l,+}}  
\nonumber \\ 
&& \quad\qquad\qquad +v_{l,+}|0\rangle \otimes |1) \otimes 
|\Phi(X)\rangle_{v_{l,+}} 
\nonumber \\
|4,l \rangle& =& u_{l,-}|\uparrow \downarrow \rangle \otimes |0) 
\otimes |\Phi(X)\rangle_{u_{l,-}}  
\nonumber \\
&& \quad\qquad\qquad +v_{l,-}|0\rangle \otimes |1) \otimes 
|\Phi(X)\rangle_{v_{l,-}} 
\nonumber \\ 
|5,l \rangle& =& |\uparrow \rangle \otimes |1) \otimes 
|\Phi(X-X_0)\rangle_l 
\nonumber \\
|6,l \rangle& =& | \downarrow \rangle \otimes |1) \otimes 
|\Phi(X-X_0)\rangle_l 
\nonumber \\
|7,l \rangle& =& |\uparrow \downarrow \rangle \otimes |1) \otimes 
|\Phi(X-X_0)\rangle_l
\quad ,
\end{eqnarray}
where $|\sigma\rangle$ denotes a site occupied by an electron with 
spin $\sigma$ and $|\!\uparrow\downarrow\rangle$ a site occupied by a 
pair of electrons with spin up and down. $|0)$ and  $|1)$ denote a site 
unoccupied and, respectively, occupied by a Boson. $|\Phi(X)\rangle_l$ 
denotes the $l$-th excited oscillator state and 
$|\Phi(X-X_0)\rangle_l= (a^+-\alpha)^l/\sqrt{l!}
\,exp(\alpha(a-a^+))|\Phi(x)\rangle_0$ the $l$-th 
excited shifted oscillator state. These two sets of oscillator states 
are sufficient to describe all the states listed in Eq.(3) 
except for the states $|3,l \rangle$ and  $|4,l \rangle$ 
for which the corresponding 
oscillator states are given by $|\Phi(X)\rangle_{u_{l,\pm}}$ and 
$|\Phi(X)\rangle_{v_{l,\pm}}$.
The latter are determined 
by numerical diagonalization by expanding them 
in a set of excited harmonic oscillator states in the form
$ u_{l,\pm}|\Phi(X)\rangle_{u_{l,\pm}}=
\sum_n u_{l,\pm}^n|\Phi(X)\rangle_n$ and
$ v_{l,\pm}|\Phi(X)\rangle_{v_{l,\pm}}=
\sum_n v_{l,\pm}^n|\Phi(X)\rangle_n$.
For the regime of coupling parameters which we are interested in we 
take into account up to $50$ phonon states, i.e., $n \leq 50$.
It is the states $|3,l \rangle$ and $|4,l \rangle$ 
which describe the transfer of polaronic features from the 
localized bipolarons to the conduction electrons when Boson-Fermion 
exchange processes take place. Since photoemission only couples to  
the electrons, it is via this transfer of polaronic features to the 
intrinsically non-polaronic electrons that photoemission spectra show  
features which are reminiscent of polaronic 
quasi-particles. These temporarily fluctuating local lattice 
deformations described by the corresponding oscillator 
wave functions $|\Phi(X)\rangle_{u_{l,\pm}}$ and  
$|\Phi(X)\rangle_{v_{l,\pm}}$ are manifest in the 
pair distribution function ($PDF$)
\begin{equation} 
g(x)=\frac{1}{Z}\sum_{n,l}\exp(-\beta E(n,l))\langle n,l| 
\delta(x) |n,l \rangle \quad . 
\end{equation}
Here $Z=\sum_{m=0}^7 \sum_{l=0}^\infty e^{-\beta E(m,l)}$ denotes 
the partition function, with $E(m,l)$ being the eigenvalues of the 
eigenstates listed above, given by:
$E(0,l)=l \hbar \omega_0$, $E(1,l)=E(2,l)=\varepsilon_0+l\hbar\omega_0$,
$E(3,l)=\varepsilon_{l,+}$, $E(4,l)=\varepsilon_{l,-}$, $E(5,l)=E(6,l)=
\varepsilon_0+E_0+l\hbar\omega_0-\varepsilon_{BP}$ and  
$E(7,l)=2\varepsilon_0+E_0+l\hbar\omega_0-\varepsilon_{BP}$, 
with $\varepsilon_0=D-\mu$ and $E_0=\Delta_B-2\mu$.

In order to investigate the various physical quantities on the basis 
of this single-site generalized $BFM$ we must choose $\Delta_B$ 
in a way to guarantee the conditions set out above, 
that is, a concentration of electrons $n_F \simeq 0.75$ 
(corresponding to a hole concentration of $\simeq 0.25$) for a total 
concentration of particles $n_{tot}=2$.
In order to achieve these conditions we put 
the bare bipolaronic level $\Delta_B$ above the bare
electronic energy level $D$ such that the bipolaronic 
levelshift $\varepsilon_{BP}$ brings this level down slightly below the 
bare electronic level. We adjust the precise position of this level by 
putting $\Delta_B=2D+\hbar \omega_0 \alpha^2-\delta\Delta_B$, with 
$\delta\Delta_B=0.025$, chosen in order to give $n_F \simeq 0.75$.
\begin{figure}
\centerline{\epsfxsize=8cm \epsfbox{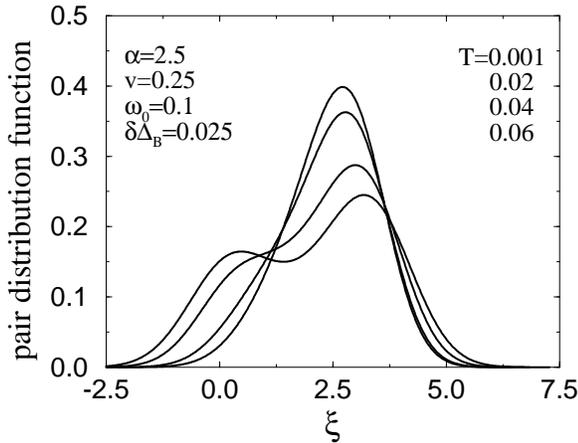}}
\caption{The pair distribution function for several temperatures $T$
in units of $D$, the highest $T$ corresponding to the most pronounced  
double-peak structure. The displacement is measured  by the dimensionless 
parameter $\xi=X \protect\sqrt{M \omega_0/\hbar}$. }
\label{figpdf}
\end{figure}
Given this choice of parameters, we obtain a $PDF$  (illustrated in 
Fig.1) showing a double-peak structure which merges into a single-peak 
structure as the temperature is lowered below the characteristic 
temperature $T^*$, at which, as we shall see below, 
the pseudogap opens up.   
The two peak positions characterize the two deformations of the local 
lattice environment where a given site is alternatively occupied by 
a pair of electrons or by a bipolaron. 
Recent $EXAFS$\cite{Roehler-97}, $XANES$\cite{Conradson-97} and 
pulsed neutron scattering\cite{Egami-96} experiments give some 
indications for such dynamical local lattice fluctuations.

Let us now embark on the evaluation of the intensity of the 
photoemission spectrum $I_{PES}(\omega)$ from 
a single site Boson-Fermion system
- tantamount to the angle integrated rather than angle resolved 
photoemission spectroscopy when neglecting the effect of the dynamical 
mean field coming from the itinerancy of the electrons.
We have $I_{PES}(\omega)=I_E(\omega)n_F(\omega)$,  
where $n_F(\omega)$ denotes the Fermi distribution function and 
$I_E(\omega)$ 
the emission part of the total one-particle Fermionic spectral function 
\begin{eqnarray}
&&I(\omega) = \frac{1}{Z}\sum_{m,m';l,l'}\left( e^{-\beta E(m,l)}+ 
e^{-\beta E(m',l')} \right) \nonumber \\
&&\qquad\quad | \langle m',l'|c_{\uparrow}|m,l \rangle |^2 
\delta(\omega-E(m,l)+E(m',l'))
\nonumber \\
&&\quad\quad = Z_F\delta(\omega-\varepsilon_0)+\frac{1}{Z}\sum_{l,m,s=\pm}
|u_{l,s}^m|^2(e^{-\beta(\varepsilon_0+m\hbar\omega_0)} \nonumber \\
&&\qquad\quad + e^{-\beta\varepsilon_{l,s}} ) 
\delta(\omega+\varepsilon_0+m\hbar\omega_0-\varepsilon_{l,s}) 
\nonumber \\
&&+{e^{-\alpha^2}\over Z} \sum_{l,m,s=\pm} \left| \sum_{n \leq l} 
v_{m,s}^n 
\sqrt{{l! \over n!}}\sum_{n'=0}^n {n \choose n'} 
{\alpha^{n-n'}(-\alpha)^{l-n'} \over(l-n'!)} \right| ^2 \nonumber \\
&&(e^{-\beta \varepsilon_{m,s}}+
e^{-\beta(\varepsilon_0+E_0+l\hbar\omega_0-\varepsilon_{BP})})
\delta(\omega+\varepsilon_{m,s}-\varepsilon_0 -E_0 
\nonumber \\
&& -l\hbar\omega_0+\varepsilon_{BP})\quad.
\end{eqnarray}
Here $Z_F=\frac{1}{Z} (1+e^{-\beta \varepsilon_0})(1+e^{-\beta  
(\varepsilon_0+E_0-\varepsilon_{BP})}) 
n_B(\hbar \omega_0)$ represents the spectral weight of the 
non-bonding contributions which accounts for the coherent part of the 
photoemission spectrum, unaffected by any coupling to the Bosons 
and hence to the phonons ($n_B(\omega)$ denotes the Bose distribution 
function). The second and third contribution to the spectral function 
$I(\omega)$ account for the incoherent part of the spectrum.
We illustrate in Fig.2 the photoemission spectral intensity 
$I_{PES}(\omega)$ for different temperatures (in units 
of $D$). For high temperatures ($T \simeq 0.06$) 
we observe a very much broadened spectral function which in shape 
comes close to that of a typical Fermi liquid. Upon lowering the 
temperature this spectral function  starts exhibiting a pseudogap 
and at the same time a broad incoherent contribution (coming from the 
second and third term of the expression for $I(\omega)$ in Eq.(5)) 
emerges. 
The incoherent part of the spectrum 
extends over a region in energy which is of the order of the  
half band width ($\simeq 0.5 eV$) and is practically 
temperature independent at low temperatures, which seems 
to be confirmed experimentally\cite{Norman-97}. The closing up of 
the pseudogap (measured as the difference in energy between the 
chemical potential at $\omega=0$ and the midpoint of the leading edge 
of the photoemission spectrum) as we increase the temperature is 
illustrated in the inset of Fig.2. 
The pseudogap has a zero temperature limit of 
$0.085D\simeq 40\,meV$ and closes up at 
a characteristic temperature $T^* \simeq 0.06D\simeq 350\,K$, which are  
reasonable numbers. We should add that the chemical potential for 
temperatures below $T^*$ turns out to be practically temperature 
independent. In order to illustrate the closing up of the pseudogap 
as the temperature approaches $T^*$ we plot in Fig.3 the density of 
states $I(\omega)$ for different temperatures. We clearly notice a 
strongly non-symmetric bias near the chemical potential ($\omega = 0$) 
which seems to be verified in tunneling experiments\cite{Renner-98}.
\begin{figure}
\centerline{\epsfxsize=8cm \epsfbox{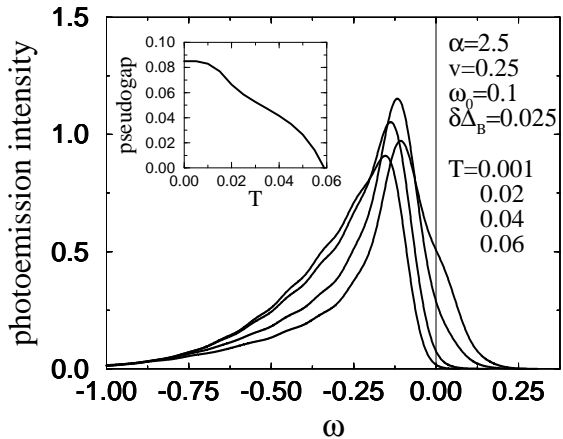}}
\caption{The photoemission spectral intensity $I_{PES}(\omega)$,
taking into account experimental broadening of $\Delta\omega=0.05$, 
for various temperatures $T$, 
the lowest $T$ corresponding to the curves with the largest pseudogap. 
The inset shows the temperature behaviour of the pseudogap.} 
\label{figsd}
\end{figure}

The work reported in this Letter relates
the temperature dependence 
of the pseudogap to that of the incoherent part of the 
quasi-particle spectrum. The opening of the pseudogap, being 
associated 
to resonant exchange tunneling between intrinsically unpaired  
electrons and electrons paired up in bipolaronic states,  
is thus driven by a metal-insulator cross-over rather than by  
superconducting fluctuations and thus can open up well 
above the onset of the superconducting phase.
The broad incoherent part of this spectrum 
is attributed to phonon shake-off effects 
arising from the polaronic character of the electrons 
in the metallic layers, transmitted to them 
via their resonant scattering into bipolaronic states. The   
temporarily fluctuating local lattice deformations which are 
caused in this process are expected to show 
up in a characteristic double-peak feature of the pair distribution 
function (measureable by $EXAFS$) and should  
test whether the incoherent $ARPES$ background is of 
polaronic origin or not. Our approach is based on an atomic 
limit calculation of the 
generalized Boson-Fermion model which is solved exactly by 
numerical means. The results obtained are not expected to change 
qualitatively when taking into account the itinerancy of the 
electrons. This will only introduce possible asymmetries 
in the Brillouin zone coming from asymmetric coupling $v$ in 
a more microscopic model (in accordance with the hypothesis of
strongly hybridized 
plane and out of plane states in certain parts of the Brilloun 
zone, as suggested by $LDA$ calculations\cite{Andersen-94}) and 
affect the quasi-particle structure close to the Fermi energy. 
For this energy regime our self-consistent 
studies\cite{Ranninger-95} on the original $BFM$, 
fully taking into account  electron itinerancy 
but neglecting any coupling to the phonons,
reproduces more faithfully the quasi particle structure,
but with an incoherent contribution which 
is typically of the order of $0.1 \; D$.  This is one order of 
magnitude less than the incoherent contributions due to phonon 
shake-off reported in this Letter. We therefore can safely treat 
this problem within the approximation scheme presented here.
Our results are moreover robust in the sense 
that they hold for different total concentrations between $1.5$ 
and $2$, as long as we enforce the condition that 
$n_F$ is between $0.7$ and 1.  
\begin{figure}
\centerline{\epsfxsize=8cm \epsfbox{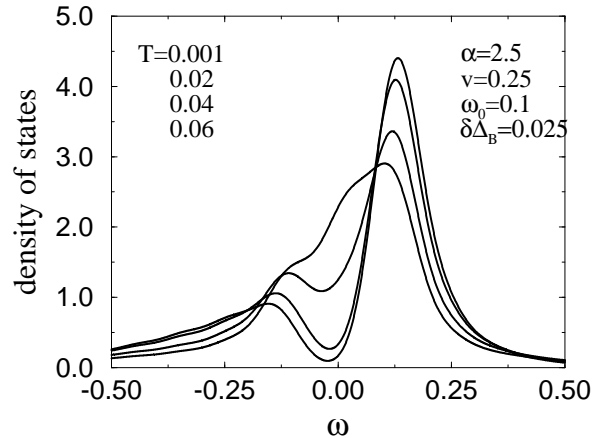}}
\caption{Evolution of the electron $DOS$ 
$I(\omega)$ as a function of temperature $T$, the lowest $T$ 
corresponding to the deepest pseudogap (experimental broadening 
$\Delta\omega=0.05$).}
\label{figdos}
\end{figure}

\end{multicols}

\end{document}